\documentclass[conference]{IEEEtran} %

\usepackage{graphicx} % Required for inserting images
\usepackage{xcolor}
\usepackage[compatibility=false]{caption}
\usepackage{subcaption} %
\usepackage{adjustbox} %
\usepackage{listings}
\usepackage{hyperref}
\usepackage{wrapfig}
\usepackage{tikz}

\usepackage[T1]{fontenc}
\usepackage{textcomp}

\usepackage{amsmath}
\usepackage{booktabs}
\usepackage{amssymb}
\usepackage{comment}

\usepackage{balance}

\definecolor{codegreen}{rgb}{0,0.6,0}
\definecolor{codegray}{rgb}{0.5,0.5,0.5}
\definecolor{codepurple}{rgb}{0.58,0,0.82}
\definecolor{backcolour}{rgb}{0.95,0.95,0.92}
\definecolor{g}{gray}{0.95}

\lstdefinestyle{mystyle}{
	backgroundcolor=\color{white},
	keywordstyle=\color{codegreen},
	numberstyle=\tiny\color{codegray},
	stringstyle=\color{codepurple},
	basicstyle=\ttfamily,
	breakatwhitespace=false,
	breaklines=true,
	captionpos=b,
	keepspaces=true,
	numbers=left,
	numbersep=5pt,
	showspaces=false,
	showstringspaces=false,
	showtabs=false,
	tabsize=2,
    morekeywords=[1]{
    },
}

\lstdefinestyle{block}{
    backgroundcolor=\color{g},
    frame=single, 
    framerule=0.5pt, 
    rulecolor=\color{gray!40},
    numberstyle=\tiny\color{gray},
    numbers=left,
    xleftmargin=15pt,
    framexleftmargin=0pt,
    numbersep=10pt,
    fillcolor=\color{g},
    identifierstyle=\color{black},
}

\makeatletter

\lstdefinelanguage{mlir}{
  morecomment = [l]{//},
  morestring=[b]", 
  sensitive = true,
  classoffset=0,
  classoffset=1, keywordstyle=\color{purple},
  morekeywords={
    soda.launch_func,
    soda.module,
    soda.return,
    func,
    define, declare, global, constant,
    internal, external, private,
    linkonce, linkonce_odr, weak, weak_odr, appending,
    common, extern_weak,
    thread_local, dllimport, dllexport,
    hidden, protected, default,
    except, deplibs,
    volatile, fastcc, coldcc, cc, ccc,
    x86_stdcallcc, x86_fastcallcc,
    ptx_kernel, ptx_device,
    signext, zeroext, inreg, sret, nounwind, noreturn,
    nocapture, byval, nest, readnone, readonly, noalias, uwtable,
    inlinehint, noinline, alwaysinline, optsize, ssp, sspreq,
    noredzone, noimplicitfloat, naked, alignstack,
    module, asm, align, tail, to,
    addrspace, section, alias, sideeffect, c, gc,
    target, datalayout, triple,
    blockaddress,
    return,
    step,
    func.func, scf.for, 
    linalg.matmul, linalg.generic,
    memref.load, memref.store, memref.subview,
    addi, muli, addf, mulf
  },
  classoffset=2, keywordstyle=\color{gray},
  morekeywords={
    memref, index, f32,
    !iarr_t, !varr_t,
    !AH_t, !AHW_t,
    !mr4x4_0, !mr4x4_1
  },
  alsoletter={\%, ., \!},
  keywordsprefix={\%},
}
\makeatother %

\usetikzlibrary{arrows.meta}
\usepackage{tcolorbox}
\tcbuselibrary{listings, skins}

\date{January 2024}
\begin{document}
%\title{Zero-Copy: Going Heap-less in AXI4MLIR}
\title{Defeat the Heap: 
Zero-Copy Data Movement in AXI4MLIR}

\author{\IEEEauthorblockN{Elam Cohavi\IEEEauthorrefmark{1},
Nicolas Bohm Agostini\IEEEauthorrefmark{3},
Jude Haris\IEEEauthorrefmark{1}, \\
Antonino Tumeo\IEEEauthorrefmark{3},
David Kaeli\IEEEauthorrefmark{2},
José Cano\IEEEauthorrefmark{1}
}
\IEEEauthorblockA{
\IEEEauthorrefmark{1}\textit{University of Glasgow},
Glasgow, Scotland, UK 
\IEEEauthorrefmark{2}\textit{Northeastern University},
Boston, MA, USA \\
\IEEEauthorrefmark{3}\textit{Pacific Northwest National Laboratory},
Richland, WA, USA
}}

\maketitle

\begin{abstract}

As custom hardware accelerators become increasingly central to machine learning workloads, efficient data transfer is critical for maximizing accelerator performance on linear algebra kernels. 
\textbf{AXI4MLIR}, an extension of the Multi-Level Intermediate Representation (MLIR) compiler framework for automated generation of host-accelerator driver code, incurs significant runtime overhead due to non-zero-copy CPU-accelerator data movement.
During transfers from the host to the accelerator, data is copied from heap-allocated memory buffers into contiguous Direct Memory Access (DMA)-mapped buffers. 
This work identifies this copy as a redundant staging operation and eliminates it through zero-copy data movement.
The optimization extends \texttt{accel}, an MLIR dialect introduced by AXI4MLIR, and implements lowering support that allocates buffers directly within DMA-mapped memory, thereby omitting the staging copy.
We evaluate the proposed scheme using a configurable matrix-matrix multiplication accelerator and show that the zero-copy optimization reduces main memory data movement by up to 2$\times$, increasing overall accelerator utilization.

\end{abstract}

%tk=16: 2.5 baseline

%tk=256: 1.09375 baseline

% optimized: 1.0078125 GB > always the same for all tks, not sure why

\section{Introduction}

As machine learning (ML) workloads continue to scale, hardware accelerators have become essential for delivering the throughput required by the linear algebra-intensive kernels prevalent in ML workloads~\cite{shabani2023hpca,kim2023hpca,hsia2023asplos,munoz2023asplos}.
% As machine learning (ML) workloads continue to scale, hardware accelerators have become essential for delivering the throughput required by linear algebra-intensive kernels that exist within ML workloads~\cite{shabani2023hpca,kim2023hpca,hsia2023asplos,munoz2023asplos}. 
For heterogeneous systems, accelerator performance is hindered by the cost of memory-based data movement arising from transfers between the host and accelerator. 

\textbf{AXI4MLIR}~\cite{agostini2023axi4mlir}, an extension of the Multi-Level Intermediate Representation (MLIR)~\cite{Lattner2021mlir} framework for describing target accelerator capabilities with arbitrary instructions, generates host-side driver code that offloads linear algebra operations to custom accelerators based on the Advanced eXtensible Interface (AXI)-Stream protocol.
AXI4MLIR operates by taking a high-level application description in MLIR’s \lstinline{linalg} abstraction~\cite{mlir2020linalg} and introducing custom MLIR attributes to specify target accelerator capabilities.
These attributes are lowered into Direct Memory Access (DMA) library calls that handle data transfers and accelerator invocation.
This design improves accelerator development effort and reduces the likelihood of errors associated with manual implementation.
However, the current lowering strategy introduces a non-trivial runtime overhead due to an additional step in the host-driver stack, with data-movement costs growing with both tensor size and tile granularity~\cite{haris2024datatransferopt}.
In particular, prior to accelerator invocation, tensors stored in standard heap-allocated memory, as \lstinline{memref} buffers, are copied into a staging buffer.
This additional data-movement overhead leads to suboptimal utilization of accelerator compute resources.

% Contributions: 
%How to mitigate the extra copy in axi4mlir? 
%What the impact of mitigating the extra copy in axi4mlir?
This work demonstrates that eliminating the staging copy between host heap memory and DMA-visible memory improves end-to-end runtime performance.
%This work demonstrates that higher accelerator utilization and, consequently, faster runtime are achieved once this extra copy is eliminated.
We introduce a zero-copy data-movement optimization that enables MLIR-allocated buffers to reside directly within DMA-mapped memory.
Our approach extends the \lstinline{accel.send} operation with a custom \lstinline{memref}-based allocation mechanism and a set of MLIR pipeline transformations that ensure accelerator-bound buffers are allocated directly in device-compatible regions. 
Eliminating this intermediary copy reduces transfer latency and host-side memory pressure. 

We evaluate the optimization on a matrix-matrix multiplication (MatMul) accelerator integrated into AXI4MLIR.
By materializing data in place in DMA-mapped memory, the proposed approach improves accelerator load/store efficiency (up to $2\times$) and increases overall compute utilization. 
These results highlight the compiler’s role in managing memory placement for heterogeneous systems and demonstrate a practical path toward reducing data-movement overhead in MLIR-based accelerator toolchains. 
Hence, the contributions of this work are as follows:

\begin{itemize}
    \item Identifying redundant data-staging overhead in host-accelerator transfer operations;

    \item Introducing a zero-copy mechanism that eliminates the intermediate staging step by allocating buffers directly within DMA-mapped memory regions;

    \item Providing a generalizable framework for integrating zero-copy semantics into MLIR-based accelerator toolchains, along with an extension plan for the MLIR \lstinline{accel} dialect to support custom memory allocation semantics.
\end{itemize}

%The proposed optimization provides a template for incorporating zero-copy semantics into other MLIR-based accelerator tooling.

% \usetikzlibrary{shapes.geometric, arrows.meta, positioning, calc}

% \begin{tikzpicture}[
%     node distance=1.5cm and 2cm,
%     block/.style={rectangle, draw, fill=grey!10, text width=5cm, align=left, rounded corners, minimum height=1em, font=\ttfamily\small},
%     arrow/.style={-Stealth, thick}
% ]

% % Nodes
% \node (subview) [block] {
%     \textbf{\%sA = memref.subview \%A} \\
%     Indices: [\%i, \%j] \\
%     Sizes: [\%tile\_size\_I, \%tile\_size\_J]
% };

% \node (copy) [block, below=of subview] {
%     \textbf{call @copy\_to\_dma\_region} \\
%     Source: \%sA \\
%     Offset: \%c4
% };

% \node (dma) [block, below=of copy] {
%     \textbf{call @dma\_start\_send} \\
%     Size: \%tile\_size\_I * \%tile\_size\_J * 4 \\
%     Offset: \%c4
% };

% \node (wait) [block, below=of dma, fill=orange!10] {
%     \textbf{call @dma\_wait\_send\_completion()} \\
%     \textit{(Synchronization Barrier)}
% };

% % Labels for IR Sources
% \node[left=0.5cm of subview, font=\scriptsize\itshape, color=gray] {Source 2};
% \node[left=0.5cm of copy, font=\scriptsize\itshape, color=gray] {Source 4};
% \node[left=0.5cm of dma, font=\scriptsize\itshape, color=gray] {Source 5};

% % Connectors
% \draw [arrow] (subview) -- (copy);
% \draw [arrow] (copy) -- (dma);
% \draw [arrow] (dma) -- (wait);

% % Data Flow Annotation
% \draw [dashed, ->] ($(copy.east)+(0,0)$)-| ++(1, -1.5) node[right, text width=2cm, font=\scriptsize] {Data moves to Accelerator DMA buffer} |- (dma.east);

% \end{tikzpicture}

\section{Background \& Motivation}

\subsection{AXI4MLIR dialect lowering}

% AXI4MLIR adopts a lowering and transformation pipeline, beginning with high-level passes on linalg-on-tensors for transformations of \lstinline{linalg} named ops (e.g., \lstinline{linalg.matmul} or \lstinline{linalg.conv_2d}). A conversion occurs into \lstinline{linalg.generic} operations based on traits, such as indexing maps and iterator types, and suitable ops are identified and annotated with custom attributes. These operations are then lowered into the \lstinline{accel} dialect, and tiling transformations are performed with new annotated attributes and the \lstinline{scf} (Structured Control Flow) dialect operation manipulation. Following a bufferization pass, \lstinline{accel} operations lower into specific runtime library calls targeting the AXI DMA engine.

AXI4MLIR adopts a multi-step lowering and transformation pipeline that begins with high-level operations on the linalg-on-tensors representation. At this stage, transformations are applied to \lstinline{linalg} named operations, such as \lstinline{linalg.matmul} and \lstinline{linalg.conv_2d}. These operations are subsequently converted into \lstinline{linalg.generic} operations based on their traits, including indexing maps and iterator types. Tensors are bufferized and represented by \lstinline{memref} operations. The converted operations are annotated with custom attributes for accelerator targeting. The annotated operations are lowered to the \lstinline{accel} dialect and other MLIR dialects, where tiling transformations are performed through a combination of \lstinline{memref} operations and manipulations of \lstinline{scf} (Structured Control Flow) dialect operations. Finaly, \lstinline{accel} operations representing host-accelerator data transfers are further lowered into runtime library calls that control the AXI DMA engine.

\subsection{AXI4MLIR host-accelerator movement}

The lowering pipeline of the baseline AXI4MLIR implementation requires explicit management that copies \lstinline{memrefs} to the DMA-visible region by producing calls to the AXI4MLIR's runtime library, such as \lstinline{@copy_to_dma_region}. Additionally, accelerator-to-host transfers follow a similar strategy, but in reverse, copying data from the accelerator to a DMA-visible host buffer and then to the destination \lstinline{memref}, which incurs a staging copy on the receive path.
%Additionally, accelerator-to-host transfers follow a similar mechanism in reverse, which incurs a staging copy on the receive path.

% \begin{figure}[t]
% \centering
% \begin{adjustbox}{minipage=\columnwidth,scale=0.95}
% %
% \centering
% \begin{subfigure}{1\textwidth}
% \centering
% \input{lsts/baseline-mlir}
% \caption{Explicit tile (\%sA) copy of the original memref into the DMA region, aligned 4 bytes from its start. }
% \end{subfigure}
% %
% \begin{subfigure}{1\textwidth}
% \centering
% \input{lsts/baseline-lowered}
% \caption{Lowered IR using calls to the DMA runtime. Baseline lowering flow has to explicitly manage copies to the DMA region.}
% \end{subfigure}
% %
% \end{adjustbox}
% \caption{Baseline lowering for accelerator transfers.}
% \label{fig:code-baseline}
% \end{figure}

Benchmarking the baseline AXI4MLIR flow for tiled matrix-matrix multiplication operations across varying problem sizes and accelerator tile dimensions reveals inefficiencies in data movement, as shown in Figure~\ref{fig:normalized_breakdown}. Each configuration uses a C-stationary algorithm~\cite{chen2016eyeriss} (the output consistist of partial results are accumulated inside the accelerator) in which $A\in\mathbb{R}^{M\times K}$, $B\in\mathbb{R}^{K\times N}$, and $C\in\mathbb{R}^{M\times N}$ are partitioned into tiles of size $T_m\times T_k$, $T_k\times T_n$, and $T_m\times T_n$. The $C$ tile remains on-chip for the duration of the inner $k$-loop, while $A$ and $B$ tiles are streamed through the accelerator. 
Problem configurations follow the format \texttt{M\_N\_K\_tile-dim-size} and span a range of accelerator sizes (e.g., \texttt{16\_16\_16\_4} through \texttt{128\_128\_128\_16}), as selected in AXI4MLIR ~\cite{agostini2023axi4mlir} micro-benchmarks.

For each micro-benchmark, we measure the fraction of total runtime spent in communication and accelerator compute, and report the baseline cases under the ``naive'' label in Figure~\ref{fig:normalized_breakdown}. The stacked bar chart shows that data movement dominates runtime under the current two-stage buffering mechanism, especially when accelerator tile sizes are small, where, for some problem configurations, the transfer cost exceeds the accelerator’s compute time by over $18\times$. This behavior holds across problem sizes and tile configurations, demonstrating that the baseline AXI4MLIR flow is more constrained by host-accelerator communication than by on-chip computation. The remainder of this work proposes an optimization to mitigate this communication cost.

\section{DMA Optimizations}

The following data movement optimizations extend the AXI4MLIR transformation and lowering pipeline to mitigate time spent on host-accelerator data move operations.

\subsection{DMA-based data allocation}

\begin{figure}[t]
  \centering
\includegraphics[width=0.95\columnwidth]{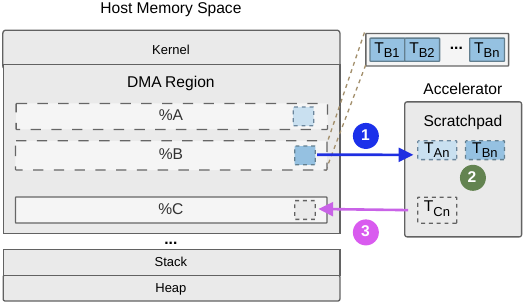}
\caption{Zero-copy DMA-aware optimization: System view.}
\label{fig:dma_zero_copy_join}
\end{figure}

To remove the staging copies present in the baseline AXI4MLIR flow, we introduce a new attribute and lowering path that allocate accelerator-visible buffers directly in the DMA memory-mapped region. Thus, DMA-to-accelerator execution consists of three steps (Figure~\ref{fig:dma_zero_copy_join}): (1) runtime-computed tiles are moved into the accelerator scratchpad for both \lstinline{%A}, \lstinline{%B} \lstinline{memref} buffers; (2) tiles (blocks) are used to compute a Matmul within the accelerator; (3) the accumulated result, tile $T_{Cn}$, is moved into its corresponding \lstinline{%C} buffer in DMA-visible region. Buffers \lstinline{%A} and \lstinline{%B} are subsequently freed.

\subsection{Reworking accel}

This optimization updates the underlying functionality of several \lstinline{accel} operations. First, \lstinline{accel.alloc} now models allocations based on an explicitly defined memory-space-type attribute, and an accompanying layout operand. For any \lstinline{accel.alloc}, the compiler emits a call to \lstinline{aximlir_dma_alloc()}, a custom allocator call which returns \lstinline{memref} descriptors allocated in DMA-visible region, capturing the physical base address, buffer size, and layout information. In MLIR, a \lstinline{memref} descriptor (Figure~\ref{fig:memref}) is composed of the necessary metadata to interpret the raw memory buffer.

\begin{figure}[hp]
\centering
\begin{adjustbox}{minipage=\columnwidth,scale=0.95}
\input{lsts/memref}
\end{adjustbox}
\caption{MLIR memref descriptor structure.}
\label{fig:memref}
\end{figure}

A \lstinline{memref.subview} produces a new descriptor that reuses the same allocated and aligned pointers from the original buffer, but updates the remaining fields to express the view: the \lstinline{offset} is adjusted to the new starting element, \lstinline{sizes} describes the dimensions of the subview, and \lstinline{strides} defines the distance (in number of elements) to step through the underlying buffer along a specific dimension. All information necessary to compute transfer ranges is therefore already present in the descriptor.

Thus, the lowering approach for \lstinline{accel.alloc} denotes a direct conversion to a \lstinline{memref.alloc} with non-zero memory space. When lowering an \lstinline{accel.send} with this flow, no intermediate copy or explicit size computation is required. 
In the worst-case scenario, one DMA call is required per line of a given tile and executes a multiple DMA operations per tile. In the best-case scenario, we can use single scatter-gather operations to move entire tiles (or subviews). 
The \lstinline{memref} or \lstinline{memref.subview} operand provides the complete set of bounds, offsets, strides, and element sizes that the runtime uses to determine the correct DMA region. The lowering emits a single call to \lstinline{dma_send_descriptor()}, passing the descriptor for the tile to be transferred. The runtime extracts the required metadata and computes the physical address and transfer length internally. A completion call is inserted when needed to preserve the blocking behavior of the original operation.

\begin{figure}[htbp]
\centering
\begin{adjustbox}{minipage=\columnwidth,scale=0.95}
\centering
\begin{subfigure}{1\textwidth}
\centering
\input{lsts/baseline-mlir}
\caption{Explicit tile (\%sA) copy of the original memref into the DMA region, aligned 4 bytes from its start.}
%\vspace{0.5cm}
\end{subfigure}
\begin{subfigure}{1\textwidth}
\centering
\input{lsts/baseline-lowered}
\caption{Lowered IR using calls to the DMA runtime. Baseline lowering flow has to explicitly manage copies to the DMA region.}
%\vspace{0.5cm}
\end{subfigure}
\centering
\begin{subfigure}{1\textwidth}
\centering
\input{lsts/dma-zero-copy-alloc}
\caption{Accelerator-visible buffers are allocated directly in the DMA memory-mapped region, eliminating the need for intermediate heap-allocated \texttt{memref} staging buffers.}
%\vspace{0.5cm}
\end{subfigure}
\begin{subfigure}{1\textwidth}
\centering
\input{lsts/zerocopy-lowered}
\caption{Lowering of \texttt{accel.send} operates directly on the \texttt{memref} or \texttt{memref.subview} descriptor, avoiding heap-to-DMA copies and issuing DMA transfers based entirely on the descriptor’s offset, size, and stride metadata.}
%\vspace{0.5cm}
\end{subfigure}
\end{adjustbox}
\caption{IR comparison of explicit copy operations with optimization-based descriptor calls }
\label{fig:dma_zero_copy}
\end{figure}

The zero-copy lowering eliminates the staging copies between heap buffers and the DMA-visible region, but it does not yet address another source of overhead that arises during transfers of higher-rank \lstinline{memref.subview} values. 
When a subview does not represent a single contiguous slice in memory, the runtime must emit one DMA transaction per contiguous region. For rank-2 (or higher) tiles, this often results in multiple transfers per \lstinline{accel.send}, especially when using the default column-major data layout. 
Since the DMA cost per call overhead is non-negligible and could become prohibitive if done line-by-line, \textit{our lowering flow prioritizes scatter-gather operations}. 
% As tile sizes grow, this fragmentation becomes a significant overhead, motivating an additional optimization that can efficiently handle non-contiguous layouts.

\subsection{Data movement with scatter-gather}

Scatter-gather DMA operations allow a device to process a list of non-contiguous memory segments as a single logical transfer. In a gather operation, the accelerator gathers multiple disjoint source regions into a contiguous on-device stream. 
Conversely, a scatter operation writes back disjoint regions into separate host addresses in a single transaction. The subview \lstinline{memref} descriptor already contains the necessary offset, size, and stride information to describe these segments. 
Our optimization constructs a gather (to send data to the accelerator) or scatter (to receive data from the accelerator) list directly from this metadata and issues a single DMA request per tile or subview, regardless of its internal contiguity. This eliminates the overhead of issuing multiple transfers, reduces per-transfer latency, and aligns the host-accelerator communication cost with the logical tile granularity defined at the MLIR level.

\begin{figure*}[!htb]
\centering
\includegraphics[width=0.98\textwidth]{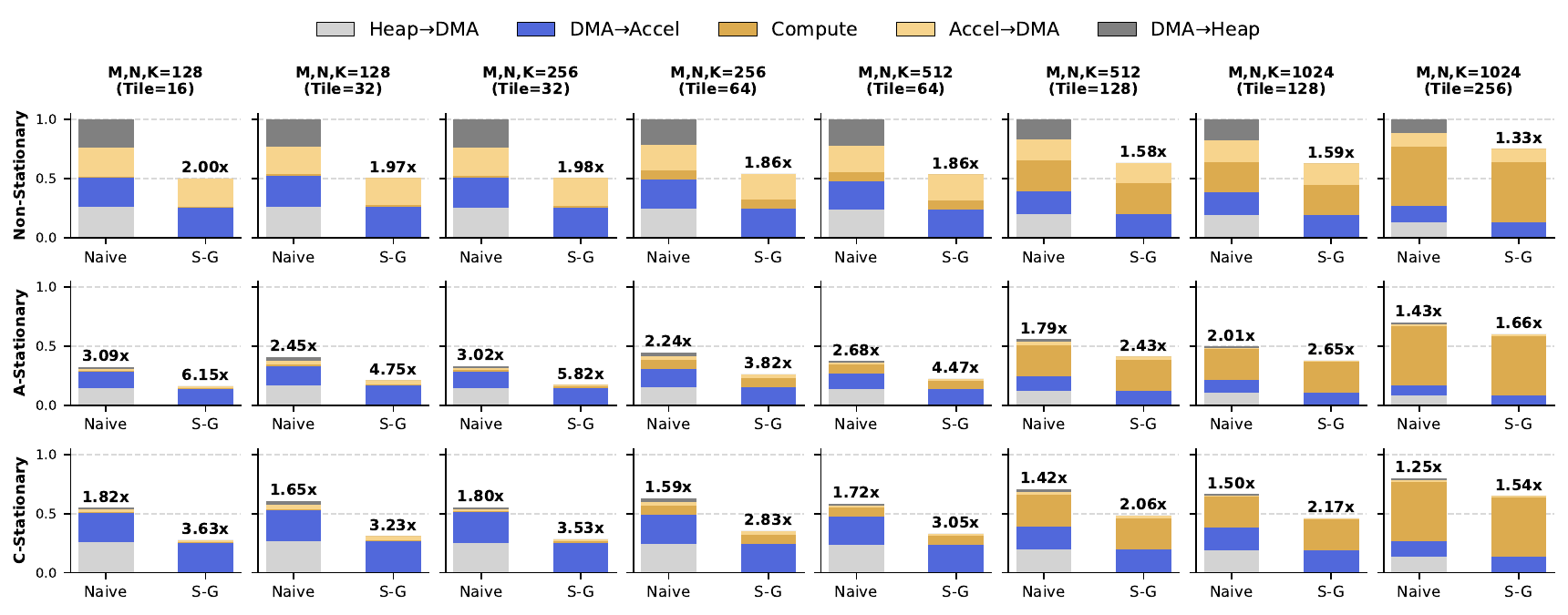}
\caption{Normalized execution time breakdown across stationary data flows for various dimensions, tile sizes.}
\label{fig:normalized_breakdown}
\end{figure*}

\section{Evaluation}

We evaluate the impact of the proposed zero-copy DMA path using a parametric data-movement model of a tiled Non-stationary, A-stationary, and C-stationary MatMul $C_{M\times N} = A_{M\times K} \times B_{K\times N}$. 
In A or C stationary flows, a given tile from A or C matrices, is read only once; but tiles associated with other matrices will be read multiple times. 
The model captures all host-device, device-host data transfers, including compute time of the tile within the accelerator. This captures two operation classes present in our runtime: 
\begin{enumerate}
  \item CPU$\rightleftarrows$DMA-region move tiles between heap to the DMA visible region;  
  \item DMA$\rightleftarrows$Accelerator transfers move data between the DMA-region and accelerator.
\end{enumerate}

For the Naive (non-zero-copy) configuration, every tile load and tile store operation induces an additional staging copy between the heap buffer and the DMA-visible region.
Our zero-copy design removes this staging overhead.

\subsection{Experimental Setup}

To evaluate the proposed optimization, we simulate a Versal VP1902 SoC characterised by a peak-aggregate memory bandwidth of 273 GB/s over a Network-on-Chip (NoC) interconnect. 
For a VP1902, the NoC is AXI4-based with a hardened (silicon-level) implementation of the AXI protocol.
The architecture leverages a low-latency NoC-based AXI traversal with a DMA latency overhead of 150ns per call, and an accelerator engine array capable of 4 TFLOPS of throughput. Simulating such parallel DSP58-based execution provides a robust environment for the following:
the implementation manages MatMul across varying problem sizes ($M=N=K \in \{128, \dots, 1024\}$) using a tiled approach where $T_M = T_N = T_K \in \{16, 32, 64, 128, 1024\}$. This ensures that a complete working set, comprised of operand tiles from $A$ and $B$ and the corresponding partial sum tile in $C$, fits within the local scratchpad memory. 

The modeled loop nest iterates over the $(M, N, K)$ index space in steps of $(T_M, T_N, T_K)$, under the stationary data flow model.
For C-stationary, the system pins the $C$ tile within local memory to maximize reuse: it is loaded exactly once, accumulated across all $k_0$ reduction steps, and moved to the host upon completion. 
In contrast, tiles of $A$ and $B$ are dynamically streamed to the accelerator as many times as needed to complete the operation.
Due to the permutation order of the C-stationary flow, $A$ tiles are reloaded at a higher frequency than $B$ tiles to maintain compute utilization. Data movement is managed via optimized DMA transfers where each $k_0$ transaction is sized according to $T_{dim} \times T_{dim}$, maximizing effective bandwidth.

\subsection{Results}

Figure~\ref{fig:normalized_breakdown} shows the normalized execution time breakdown across four stationary data flows: Non-Stationary, A-Stationary, and C-Stationary as a function of problem scale and tile size $T$. 
Each bar is normalized to the \textit{Naive Non-Stationary} baseline to highlight the relative efficiency gains. The stacked components represent the relative time spent in: heap-to-DMA, DMA-to-accelerator, Compute, accelerator-to-DMA, DMA-to-heap.
We note that evaluations for accelerators supporting a tile size of $T_K\in \{4,8\}$ are omitted because the data movement time is significantly higher than the compute time in the given setup.
As tile size $T$ increases relative to the problem dimensions, the aggregate data movement frequency (and total latency) decreases, leading to a higher proportion of time spent in the \textit{Compute} phase.
However, in the \textit{Naive} (baseline) implementations, the system incurs a significant (up to $2\times$) staging-copy overhead that scales with tile dimensions. 
The results demonstrate that Scatter-Gather (\textit{S-G}) implementations eliminate the $Heap \rightleftarrows DMA$ bottleneck by removing the need for intermediate buffer copies. In configurations with bigger tile sizes (e.g., $M,N,K=1024, T=128$), Scatter-Gather achieves a significant reduction in normalized runtime by coalescing strided memory accesses into a single DMA move operation per tile. Additionally, $A$ and $C$-Stationary flows reuse stationary tiles, further reducing off-chip traffic. 

Therefore, our model confirms the notion: as tile size increases and where staging costs would otherwise dominate, the zero-copy Scatter-Gather approach lowers memory bandwidth pressure and significantly improves ($1.7\times$ on average) end-to-end performance across all evaluated accelerator sizes and data flows.

\section{Related Work}

% \textbf{MLIR for Heterogeneous Data Movement Abstraction.} 
Hardware-software co-design enables efficient mapping of applications into custom hardware~\cite{DLAS_TACO2025}, and the use of MLIR to model and optimize data transfers between CPUs and accelerators is an emerging research area. 
Specifically, the MLIR-AIE toolchain~\cite{mliraie}, designed for AMD's Versal AI Engine (AIE) architectures, employs custom dialects to manage data flow and low-level hardware configuration.
This framework abstracts DMA-based data transfers across memory tiles and shim interfaces using explicit operations (e.g., \texttt{aiex.dma\_start\_bd\_chain} and \texttt{aiex.dma\_await\_task}).
These operations enable the compiler to statically allocate resources and orchestrate the necessary communication channels for efficient data movement across the tiled architecture, which is inherently dependent on specialized DMA engines for performance. 
While useful, this framework is for the AI Engines only and does not support custom accelerators, AXI4MLIR fills in this gap.

% \textbf{Compiler-Driven Simulation and Event Modeling.} 
For more general event-driven heterogeneous systems, the \lstinline{EQueue} dialect has been introduced to provide a foundation for compiler-driven simulation and optimization~\cite{li2022compilerdrivensimulationreconfigurablehardware}. This MLIR dialect models the asynchronous nature of hardware operations, including data movement between distinct memories and compute units. By representing control flow as events and task queues, this dialect allows the compiler to explicitly model and reason about the necessary synchronization and dependencies for efficient execution. While not exclusively a DMA abstraction, EQueue provides the essential control-flow mechanisms required to accurately model the initiation, completion, and stalls associated with asynchronous DMA transfers in a complex, heterogeneous environment.

\section{Conclusion}

This work identified and addressed inefficiency in the AXI4MLIR host-accelerator data movement pipeline: redundant staging copy between heap-allocated memrefs and DMA-mapped buffers. Using a zero-copy approach implemented through custom \texttt{memref} allocation semantics within the MLIR \texttt{accel} dialect, this enables tensors to be allocated and mapped directly into DMA-accessible regions, eliminating intermediate staging during data movement operations that accounted for a significant portion of the end-to-end latency. 
Evaluated on a tiled MatMul accelerator, the proposed optimization demonstrates substantial reductions ($1.7\times$ on average) in both load and store operation latency and further improves accelerator utilization.
The framework provides a template for integrating zero-copy data-movement strategies into other MLIR-based accelerator flows, and underscores the value of co-designing compiler and runtime abstractions to enhance performance in heterogeneous systems. 
We plan to further refine our proposed optimization by enabling custom data layouts and allocation depending on accelerator design.

\section*{Acknowledgment}

This work was partially supported by the EU Project dAIEDGE (GA Nr 101120726) and the Innovate UK Horizon Europe Guarantee (GA Nr 10090788). 
This work was partially supported by the U.S. Department of Energy (DOE) Office of Science, Office of Advanced Scientific Computing Research (ASCR), under the End-to-end co-design for performance, energy efficiency, and security in AI-enabled computational science (ENCODE) project and the Democratization of Co-design for Energy-Efficient Heterogeneous Computing (DeCoDe) project.

\balance

\bibliographystyle{IEEEtran} %
\bibliography{main}

\end{document}